\documentclass[acmsmall,nonacm]{acmart}

\usepackage{booktabs}

\setcopyright{rightsretained}
\makeatletter
\def\@acmBadgeL@image{}
\def\@acmBadgeR@image{}
\makeatother

\title{Institutions for the Post-Scarcity of Judgment}

\author{Lauri Lov\'en}
\orcid{0000-0001-9475-4839}
\email{lauri.loven@oulu.fi}
\affiliation{%
  \institution{Future Computing Group, University of Oulu}
  \city{Oulu}
  \country{Finland}
}

\begin{document}

\begin{abstract}
Each major technological revolution inverts a particular scarcity and rebuilds institutions around the shift. The near-consensus diagnosis of the AI revolution holds that AI collapses the cost of prediction while judgment remains scarce. This Opinion argues the inversion has now flipped: competent-looking judgment (selecting, ranking, attributing, certifying) is produced at scale and at marginal cost approaching zero, and four complements become scarce: verified signal, legitimacy, authentic provenance, and integration capacity (the community's tolerance for delegated cognition). Because judgment is the substance of institutions, the institutions built to manufacture legitimate judgment (courts, journals, licensing bodies, legislatures) now compete with the technology for the same functional role. The piece traces the pattern across scientific institutions, professional licensing, intellectual property, democratic legitimacy, and foundation-model concentration, and closes with a three-move agenda: reframe AI policy as institutional redesign, build provenance and verification as commons, and develop the formal apparatus for institutional composition under strategic agents.
\end{abstract}

\keywords{institutional design, AI governance, content provenance, legitimacy, commons governance, credentialing, technology policy}


\maketitle

\section{The scarcity-structure frame}

The techno-economic-paradigm tradition argues that each revolution is legible through one diagnostic: what moved from scarce to abundant, and which institutions followed. Perez~\cite{perez2002} tracks five revolutions: agriculture made food abundant and rebuilt society around land tenure; industry made mechanical power abundant and rebuilt it around capital and wage employment; information made signal abundant and rebuilt commerce around attention, data, and platforms. Brynjolfsson and McAfee~\cite{brynjolfssonmcafee2014} state the tension in one line: technology moves fast, institutions do not keep up. Which scarcity the AI revolution inverts is the question.

The near-consensus answer, popularised by Agrawal, Gans, and Goldfarb~\cite{agrawalgansgoldfarb2022}, is that AI collapses the cost of \textit{prediction} while \textit{judgment} remains the expensive complement. We argue the inversion has now flipped. When a frontier model ranks job applications, drafts a medical differential, writes a peer review, composes a legal brief, or certifies a compliance report, it is not producing raw prediction for a human to judge. It is producing competent-looking judgment, at scale, at marginal cost approaching zero. Judgment is the new abundance; the scarce complements are verification, legitimacy, and authentic provenance.

This matters because land, machinery, and signal are raw materials institutions coordinate, whereas judgment is what institutions manufacture. Courts manufacture legal judgment; universities, scholarly judgment; licensing bodies, adjudicated competence; legislatures, legitimate collective judgment. When the substance of institutions becomes abundant, they compete with the technology for the same functional role. The post-2022 policy debate, conducted in the register of harms-mitigation and enterprise adoption, is missing its own object. Ostrom's commons-design principles~\cite{ostrom1990}, extended by Hess and Ostrom to knowledge and infrastructural goods~\cite{hessostrom2007}, give the starting apparatus, not yet applied as a single integrative frame to the judgment-manufacturing institutions now under AI-era pressure.

\section{The new scarcities}

When judgment becomes abundant, four previously-free goods become scarce and load-bearing. The list is not exhaustive, but institutions cannot route around these.

\textit{Verified signal} is the first. Institutions long relied on the correlation between competent-looking output and underlying competence. AI breaks the correlation by lowering the production cost of plausibility faster than the cost of verification. A claim that has been checked, reproduced, attested, or adversarially tested carries information that a merely ``reviewed'' claim does not. Verification infrastructure, not certification branding, is the scarce good. Standards such as C2PA~\cite{c2pa} operationalise this at the artefact level; the analogous infrastructure for institutional outputs (legal opinions, audit reports, diagnostic recommendations, peer reviews) does not yet exist.

\textit{Legitimacy} is the second. Legitimacy is the social authority to call a judgment authoritative, and it behaves as a commons-like shared stock (subtractable through reputational degradation, excludable only imperfectly, so Ostrom's governance apparatus~\cite{ostrom1990} applies to replenishment even where the textbook common-pool-resource classification does not fit exactly). It replenishes when outputs track reality and depletes when marginal outputs (unattributed AI-assisted reviews, AI-drafted rulings, AI-filed submissions no human verified) are produced without replenishing the underlying practice. Institutions that overspend their legitimacy budget do not announce it; they become irrelevant to the communities they were built to serve while surface activity continues.

\textit{Authentic provenance} is the third. Which model produced a claim, which data it was trained on, which prompt elicited it, which human decisions shaped it: current metadata regimes do not carry these facts. Copyright, credentialing, and liability assume a traceable chain of human judgment. AI-augmented workflows produce artefacts whose provenance is a graph, and downstream institutions have no vocabulary for graph-valued attribution. Provenance infrastructure is a systems-research problem the systems-research community will build or fail to.

\textit{Integration capacity} is the fourth and the one with no close prior. An institution can absorb only so much delegated AI-judgment before its community stops treating its outputs as authoritative. The binding constraint is not computational throughput; it is the community's tolerance for delegated cognition. It has to be designed, not messaged, and admits operational handles: disclosure-weighted legitimacy indices, dissent rates from the served community, and opt-out uptake when a human-only track is offered. Lagging, but measurable.

The four individually have precedent. Their assembly into a single institutional-design taxonomy with engineering handles, with integration capacity as a previously-unrecognised scarcity within that set, is the contribution. Sections 3 through 7 trace the pattern across five domains.

\section{Scientific institutions as one case}

Science is the sharpest instance because its core institutions (peer review, authorship, reproducibility, grant evaluation) each encode an explicit scarcity assumption that AI flips. Peer review rations expert attention across submission volume; when submission volume scales with AI-assisted writing faster than review capacity, the process degrades from the inside while surface activity continues. Authorship norms attribute a scarce cognitive contribution to an identifiable individual and have no vocabulary for a human-AI workflow in which the intellectual substance is co-produced. Reproducibility practices rest on publication as a probabilistic signal of truth; when the cost of producing plausible but unreproducible work falls faster than the cost of verifying it, publication stops being a reliable signal.

Each failure mode is institutional, not technical, and compounds silently. Science is one instance of a family. Susskind and Susskind~\cite{susskindsusskind2022} document the same shape in law, medicine, accountancy, and consulting: gatekeeping institutions whose scarce input is being standardised at scale while institutional response lags. Mechanism differs (reputation-backed institutions deplete through signalling erosion; enforcement-backed through backlog and rule-gaming), but the shape recurs.

\section{Labour markets and professional licensing}

Susskind and Susskind~\cite{susskindsusskind2022} argued a decade ago that professional expertise would be unbundled by technology, and credentialing regimes are not obviously the right institutional form for a world in which practical expertise is standardised at scale. The AI-era specialisation is sharper: credentialing regimes certify a good (scarce human judgment) that is no longer scarce, and the institutions themselves, not only their outputs, now compete with the technology. Three consequences follow, and the current policy debate conflates them.

First, the credential still binds: you still need a licensed physician to prescribe, a licensed attorney to appear, a licensed engineer to sign off on a bridge. Second, the \textit{content} of the gated act is increasingly AI-co-produced: the differential diagnosis is triaged by a model, the contract review drafted by one, the audit opinion drawing on agentic tooling. Licensing regimes have no vocabulary for what the credential asserts when the substance of the gated output did not originate in the credentialed human. Third, the economic equilibrium breaks: price remains pinned by credential supply while marginal cost collapses, producing politically unstable, legally ambiguous rents.

The design space is richer than ``retrain displaced workers.'' Narrower credentials could certify specific AI-augmented workflows, not the whole practitioner; liability-layer credentials could make the holder strictly liable for AI-produced outputs signed; audit-of-augmentation regimes could certify the workflow. Prior art exists: aviation distinguishes type ratings from the pilot's certificate, and sell-side finance audits research workflows under MAR and MiFID. Translation to medicine, law, engineering, and accountancy has not been systematic.

The policy fight over ``AI displacing jobs'' is the wrong frame: the real question is what the credential is \textit{for} when the judgment it certifies is no longer scarce. Current legislation (the EU AI Act~\cite{euaiact2024}, the evolving US state patchwork) regulates AI systems, not the credential-judgment coupling AI is dissolving. Concrete vehicles (EU AI Act codes of practice, ISO/IEC 42001, sectoral licensing statutes) could be recast for workflow auditability rather than model behaviour. Computing researchers have a stake: we write the tooling that makes the coupling inspectable, or we do not.

\section{Intellectual property regimes}

Copyright and patent regimes were built on a composite premise: the protected expression or invention is a scarce output of scarce human judgment, and both halves held firmly enough that authorship or inventorship was a tractable abstraction. Both halves are now under pressure. The expression is cheap, and the judgment that produced it may not be human, may be distributed across human and model contributions, or may be uncertain as to which human contributed which part.

Current IP litigation (NYT v. OpenAI, the image-generator training cases, the Andersen-style suits) is fighting the last war over whether training constitutes reproduction. That is downstream of the institutional question: what is the right granularity of protection when a prompt, a fine-tuned model, a base model, a dataset, and a downstream human edit each embed partial judgment, and the output is the product of all of them?

The authorship and inventorship abstractions are the institutional debt. The replacement is a provenance-and-attribution regime at the granularity of \textit{contribution chains}, closer to open-source license propagation than to Berne Convention authorship. This does not require repealing copyright; it requires layering a provenance regime above it, tagged to artefact-level attestation of the form C2PA~\cite{c2pa} has begun to formalise for media (C2PA covers media artefacts, not multi-component institutional outputs; the latter needs an extension). The natural instrument is not a treaty or statute but a standards-body deliverable, adopted by reference in sectoral statutes and procurement contracts. Institutional question: which chain layer protection attaches to. Engineering question: how to keep the chain inspectable without destroying legitimate incidental use. Both are in computing's court.

\section{Democratic legitimacy and compute as commons}

Acemoglu and Johnson~\cite{acemoglujohnson2023} argue that technology is a political choice and that whether a revolution produces broadly-shared prosperity or elite concentration depends on institutional design and who controls the technology. Applied to AI, the default path leads to concentration and democratic erosion unless institutions force a different one. This Opinion descends one level, to the mechanism: the institutional substrate becomes contested, because judgment-abundance competes directly with the manufacture of legitimate collective judgment.

One-person-one-vote encodes a now-pressured claim: the cognitive capacity a citizen brings to a collective decision is roughly uniform, so equal weight is fair and feasible. The assumption survived the industrial age because physical capital was decoupled from cognitive capacity. Universal decision-support breaks the decoupling: cognitive capacity is now a function of which model a citizen can run, who trained it, and on whose data. Pre-existing digital asymmetries (broadband, platform concentration, language support) load onto this; concentration pressure is strongest where those asymmetries are deepest. Frontier-model access becomes a gating factor on effective political participation under conditions current democratic theory was not built to analyse.

The institutional question is whether compute and foundation-model access should be treated as public infrastructure. Following Ostrom and Hess-Ostrom~\cite{ostrom1990,hessostrom2007}, they should: foundation models and training compute constitute commons-governed infrastructure with mixed rivalrous structure (rival in production, non-rival in inference). Of Ostrom's eight principles, nested enterprises and clear boundaries map to a federation of public-compute cooperatives; principles presupposing low-cost monitoring transfer less cleanly under model opacity, itself an engineering problem the community should own.

Four counter-pressures are on the policy table. \textit{Open-weight frontier models}, necessary for auditability, insufficient because training compute is the binding constraint. \textit{Edge compute}, necessary for sovereignty and latency, insufficient because model provenance sits upstream. \textit{Public-compute cooperatives} (e.g., EuroHPC AI Factories, the U.S. National AI Research Resource), necessary for democratic footing, insufficient without governance theory to guide allocation. \textit{Sovereignty stacks} (vertically integrated national or bloc-level AI stacks such as the EU AI-gigafactory programme), necessary for political control, historically prone to capture through procurement concentration around national-champion vendors. Each relocates capture risk rather than escaping it: cooperatives shift gatekeeping from capital markets to allocation-rule design, not eliminate it.

The binding scarcity in the long run is not compute. Compute is a resource problem with a two-decade resolution horizon; legitimacy is an institutional problem that does not resolve on its own. The EU AI Act~\cite{euaiact2024} provides partial scaffolding on accountability but does not address compute-as-commons, and the deeper scarcity it does not speak to is integration capacity (section 2), which no amount of compute investment can buy.

\section{Concentration versus democratisation}

The default trajectory of foundation-model markets is concentration, driven by capital intensity, data gravity, talent agglomeration, and regulatory moats that inadvertently favour scale. The default governance response is fragmentation: the EU AI Act~\cite{euaiact2024}, a US patchwork across SB 53 and sectoral rules, China's state-centric regime, and a long tail of national strategies, producing jurisdictional arbitrage rather than democratisation on the deployment side. The mechanism-level question (by what architecture the counter-pressures compose) is where computing takes over.

The four structural correctives are individually necessary but \textit{composition-counterproductive}: partial composition can score worse than no composition, because strategic actors optimise against the implemented fraction while the absent fraction licenses the failure modes the composition was meant to prevent (a Goodhart-style dynamic where the enacted subset becomes the target). \textit{Open weights without public compute}: auditability collapses back to whoever owns the GPUs. \textit{Public compute without open weights}: a cooperative allocating access to closed frontier models relocates gatekeeping to whoever runs the cooperative. The claim is about naive pairwise implementation, not impossibility. Non-failing paths exist (cooperatives with open-weights access mandates, federated compute commons under open-weights licences, countervailing-power arrangements); the decisive variable is whether allocation rules, licences, and procurement terms make the implemented fraction inseparable from the absent one.

National AI strategies should be evaluated on how they compose with peers, not in isolation. Who governs the composition, on what authority, with what accountability, is the binding institutional-design problem.

\section{A research agenda}

Three concrete moves follow. First, stop treating AI policy as a list of harms to mitigate and start treating it as institutional redesign, with the EU AI Act implementation-review track as the immediate vehicle. Harms-mitigation is necessary but does not substitute.

Second, build provenance, verification, and legitimacy infrastructure as commons, not enterprise products, starting with a W3C-PROV-extended institutional-output manifest as a concrete pilot. C2PA-style attestation~\cite{c2pa}, reproducibility pipelines, machine-readable provenance for model-produced artefacts, and workflow-level credentialing are systems-research problems with governance consequences. Built as commons~\cite{ostrom1990,hessostrom2007}, they replenish the legitimacy they draw on; built as products, they cannot.

Third, develop the formal apparatus for institutional composition under strategic agents, with a composition-under-strategic-agents benchmark (adversarial evaluation of paired correctives) as a proximate target. Mechanism design, scoring-rule theory, and institutional economics in the North and Ostrom traditions bear directly; building the coalition (open-source advocates, public-infrastructure blocs, credential-reform associations, regulators) is a construction problem, not an obstacle. Judgment is becoming abundant; the institutions that manufacture it now compete with the technology, and the scarce good is \textit{legitimate} judgment under strategic agents.

\bibliographystyle{ACM-Reference-Format}
\bibliography{references}

@book{perez2002,
  author    = {Perez, Carlota},
  title     = {Technological Revolutions and Financial Capital: The Dynamics of Bubbles and Golden Ages},
  publisher = {Edward Elgar},
  year      = {2002}
}

@book{brynjolfssonmcafee2014,
  author    = {Brynjolfsson, Erik and McAfee, Andrew},
  title     = {The Second Machine Age: Work, Progress, and Prosperity in a Time of Brilliant Technologies},
  publisher = {W. W. Norton},
  year      = {2014}
}

@book{agrawalgansgoldfarb2022,
  author    = {Agrawal, Ajay and Gans, Joshua and Goldfarb, Avi},
  title     = {Power and Prediction: The Disruptive Economics of Artificial Intelligence},
  publisher = {Harvard Business Review Press},
  year      = {2022}
}

@book{hessostrom2007,
  editor    = {Hess, Charlotte and Ostrom, Elinor},
  title     = {Understanding Knowledge as a Commons: From Theory to Practice},
  publisher = {MIT Press},
  year      = {2007}
}

@book{ostrom1990,
  author    = {Ostrom, Elinor},
  title     = {Governing the Commons: The Evolution of Institutions for Collective Action},
  publisher = {Cambridge University Press},
  year      = {1990}
}

@misc{c2pa,
  author       = {{Coalition for Content Provenance and Authenticity}},
  title        = {{C2PA} Technical Specification, v2.3},
  howpublished = {\url{https://c2pa.org}},
  note         = {Accessed April 2026}
}

@book{susskindsusskind2022,
  author    = {Susskind, Richard and Susskind, Daniel},
  title     = {The Future of the Professions: How Technology Will Transform the Work of Human Experts},
  edition   = {Updated},
  publisher = {Oxford University Press},
  year      = {2022}
}

@misc{euaiact2024,
  author       = {{European Union}},
  title        = {Regulation ({EU}) 2024/1689 laying down harmonised rules on artificial intelligence ({Artificial Intelligence Act})},
  howpublished = {Official Journal of the European Union, OJ L, 2024/1689},
  year         = {2024},
  note         = {Published 12 July 2024; entered into force 1 August 2024}
}

@book{acemoglujohnson2023,
  author    = {Acemoglu, Daron and Johnson, Simon},
  title     = {Power and Progress: Our Thousand-Year Struggle Over Technology and Prosperity},
  publisher = {PublicAffairs},
  year      = {2023}
}

\end{document}